\documentclass[psfig]{aa} 

\usepackage{times}
\usepackage{graphics}
\usepackage{latexsym}
\usepackage{epsfig}

\begin{document}

\title{Exceptional H$_{\rm 2}$ emission in the Antennae galaxies: \\
  Pre-starburst shocks from the galaxy collision\thanks{Based on
  observations with the Infrared Space Observatory 
  ISO, an ESA project with instruments 
  funded by ESA Member States (especially the PI countries: France, 
  Germany, the Netherlands and the United Kingdom) and
  with the participation of ISAS and NASA.}
} 

\author{Martin Haas\inst{1}
   \and Rolf Chini\inst{1}
   \and Ulrich Klaas\inst{2}
}
\offprints{Martin Haas (haas@astro.rub.de)}
\institute{
Astronomisches Institut, Ruhr-Universit\"at Bochum (AIRUB),
Universit\"atsstr. 150 / NA7, 44780 Bochum, Germany
\and
Max-Planck-Institut f\"ur Astronomie, 
K\"onigstuhl 17, 69117 Heidelberg, Germany
}

\date{Received 13. Dec. 2004; accepted 20. Jan. 2005}
\authorrunning{M. Haas et al.}
\titlerunning{Pre-starburst shocks in the Antennae}

\abstract{
The collision of gas-rich galaxies is believed to produce strong
shocks between their gas clouds which cause the onset of the observed
bursts of extended star formation.
However, the so far observed shock signatures in colliding galaxies
can be explained essentially by winds from already existing massive
stars and supernovae and thus do not give any evidence for an
outstanding pre-starburst phase.
Either pre-starburst gas shocks are too short-lived to be detected
or one has to modify our perception of colliding galaxies.
A dedicated analysis of ISOCAM-CVF mid-infrared spectral maps led us
to the 
discovery of exceptional
H$_{\rm 2}$ v=0-0 S(3) $\lambda$=9.66$\mu$m line emission from
the "Antennae" galaxy pair, which is at an early stage of galaxy collision.
Its H$_{\rm 2}$ line luminosity, normalized by the far-infrared
luminosity, exceeds that of all
other known galaxies and the strongest H$_{\rm 2}$  emission is spatially
displaced from the known starbursts regions. This implies that most of
the excited H$_{\rm 2}$  gas in the Antennae must be shocked due to the
collision of the two galaxies. These observations indicate that the
outstanding phase of pre-starburst shocks exists, and that they might
be a key to our understanding of the formation of the first
proto-galaxies.
}
\maketitle

\section{Introduction} 
\label{section_introduction}

Bursts of star formation are frequently observed in gas-rich
interacting galaxies, leading to a picture of different evolutionary
starburst phases. Consensus is growing that after the first ignition
of massive stars, further cascades of starbursts are triggered by the
blast waves of massive stars and supernovae compressing the
surrounding medium, as is indicated by observations of star forming
regions in our Galaxy (\"Ogelman \& Maran 1976, Elmegreen \& Lada
1977). However, very little is known about the
conditions immediately before the onset of the explosive star
formation event, in particular the pre-starburst phase at the
beginning of a cascade. Establishing basic principles thereof would be
of general value.

Theoretical considerations as well as numerical
simulations suggest that the ignition of starbursts requires not only 
dense gas reservoirs, but also shocks causing the gas clouds to
collapse (Scoville et al. 1986, Jog \& Solomon 1992, Barnes 2004).
Although this picture is widely accepted, direct
evidence for such pre-starburst shocks is yet missing due to the
difficulty of observing the presumably short-lived phase itself.
Any so far detected shock signatures in colliding galaxies
can be explained by winds from already existing massive stars and
supernovae (e.g.
Campbell \& Willner 1989, Kunze et al. 1996, Rigopoulou et al. 2002,
Lutz et al. 2003, Ohyama et al. 2004).
Furthermore, colliding systems in an advanced merger stage
show already strong relicts from previous starbursts. Since it may be
hard to separate regions of already ongoing starbursts from those in a
pre-starburst phase, the challenge is to find a rather virgin pair of
colliding galaxies where most of the gas is still on the verge of
collapse.

NGC 4038/4039 is the prototype of a colliding galaxy pair, due to its
long tidal tails nicknamed the "Antennae".
The system is at an early stage of
encounter (Toomre 1977, Mihos \& Hernquist 1996).
Although a luminous infrared galaxy with 
L$_{\rm FIR}$$\sim$5$\times$10$^{\rm 10}$\,L$_{\odot}$ (Klaas et al. 1997), 
its current star formation efficiency is yet low,
with an average value
L$_{\rm FIR}$/M$_{\rm gas}$\,=\,4\,L$_{\odot}$/M$_{\odot}$
comparable to that of normal star forming galaxies
(Gao et al. 2001).
Of particular interest is the overlap region of the two galaxy
disks which exhibits a large amount of molecular gas
(Stanford et al. 1990, Young et al. 1995,
Wilson et al. 2000, Gao et al. 2001), permeated
by compressed magnetic fields (Chyzy \& Beck 2004).
There are ongoing starbursts, the
most violent ones are still heavily dust-enshrouded and located
south of the molecular gas concentrations (Vigroux et al. 1996,
Mirabel et al. 1998).  Hence, the
Antennae are in a stage of further imminent extra-nuclear starbursts
(Haas et al. 2000, Gao et al. 2001)
and therefore well suited to search for pre-starburst
shocks.

A well known observational signature for shocks is the line
emission of molecular hydrogen (H$_{\rm 2}$) although its excited states may
also be induced by hard photons from already existing starbursts or by
winds from supernova explosions (e.g. Hollenbach \& McKee 1989,
Sternberg \& Neufeld 1999). So far, H$_{\rm 2}$  line emission has
been detected in three small areas of the Antennae: The two nuclei
display fairly inconspicuous H$_{\rm 2}$ luminosities
compared to other starburst systems (Campbell \& Willner 1989),
while the H$_{\rm 2}$  emission in the
southern edge of the
overlap region has been attributed to the active starbursts there
(Kunze et al. 1996).
Hence, pre-starburst shocks have to be searched for in other
regions of the Antennae.

\section{Data}
\label{section_data}

The ISO Data Archive (Kessler et al. 2003) provides valuable
mid-infrared spectral maps
of the entire Antennae system obtained with the ISOCAM Circular
Variable Filter (CVF) mode (Cesarsky et al. 1996).
Since only small portions of this data set have
yet been published addressing other topics (Vigroux et al. 1996,
Mirabel et al. 1998, Haas et al. 2002), we have
reduced and evaluated the full 3D data cube. 
We checked the CVF frames and photometry 
by comparing with
images in the 6.0, 6.7, 9.6, and 14.3 $\mu$m filters.
By visual inspection we assured that the CVF frames
do neither show ghost features, which may sometimes occur, nor any 
measurable effect of straylight within the
typical calibration accuracy of about 30\% (Blommaert et al. 2003).  

\section{Results and Discussion}
\label{section_results}
Figure 1  shows the 5-16$\mu$m
spectrum derived from a region, which encompasses the two nuclei as
well as the overlap region in-between. Among several spectral features
we focus here on the well discerned rotational H$_{\rm 2}$  line at
$\lambda$$_{\rm obs}$ = 9.72
$\mu$m ($\lambda$$_{\rm rest}$ = 9.66 $\mu$m),
designated in detail H$_{\rm 2}$  v = 0-0 S(3).

Figure 2 
shows the total continuum-subtracted H$_{\rm 2}$  line map superimposed on an
optical three colour image of the Antennae. Obviously, most of the H$_{\rm 2}$ 
line emission arises from the overlap region.
While the most active starbursts are concentrated in the
southern part of the overlap region (Mirabel et al. 1998)
the northern part is less
active displaying five times weaker starburst ionisation lines
(Vigroux  et al. 1996),
cooler dust (Haas et al. 2000),
more regular and stronger compressed magnetic fields
(Chyzy \& Beck 2004) and ten times fainter X-ray emission on the
Chandra map (Fabbiano et al. 2000). In contrast, the
H$_{\rm 2}$  line emission is evenly strong in both the southern and the
northern part of the overlap region suggesting that most of the H$_{\rm 2}$ 
line emission is generated independently of the already active
starbursts.

\begin{figure}
  \begin{center}
  \epsfig{file=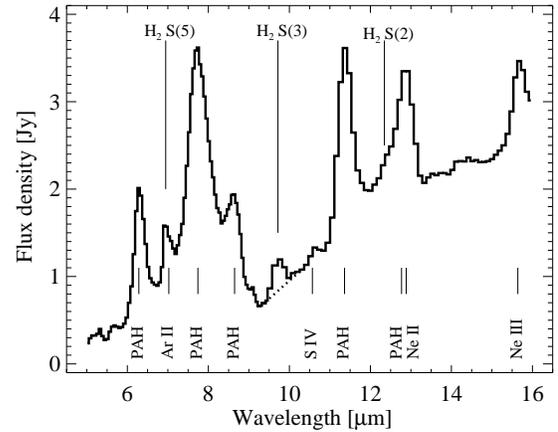 ,width=8.5cm, clip=true}
  \caption[]{ 
    Mid-infrared spectrum of the Antennae, obtained with the
    ISOCAM-CVF mode. It has been derived from a 
    2'$\times$2' region,
    encompassing both nuclei and the entire overlap region. The
    prominent emission features are those of polycyclic aromatic
    hydro-carbons (PAHs) and Ne lines.
    Three H$_{\rm 2}$ 
    lines are marked.
    While the H$_{\rm 2}$  S(5) and H$_{\rm 2}$  S(2) lines are blended with
    [ArII] $\lambda$ = 6.99 $\mu$m and [NeII] $\lambda$ = 12.8 $\mu$m,
    respectively, the H$_{\rm 2}$  S(3)
    line can unambiguously be detected above the continuum (dotted). The
    zodiacal and galactic foreground spectrum has been subtracted as derived from the
    36" wide border of the 192"$\times$192" spectral maps.  
    }
  \end{center}
\end{figure}
\begin{figure*}
    \epsfig{file=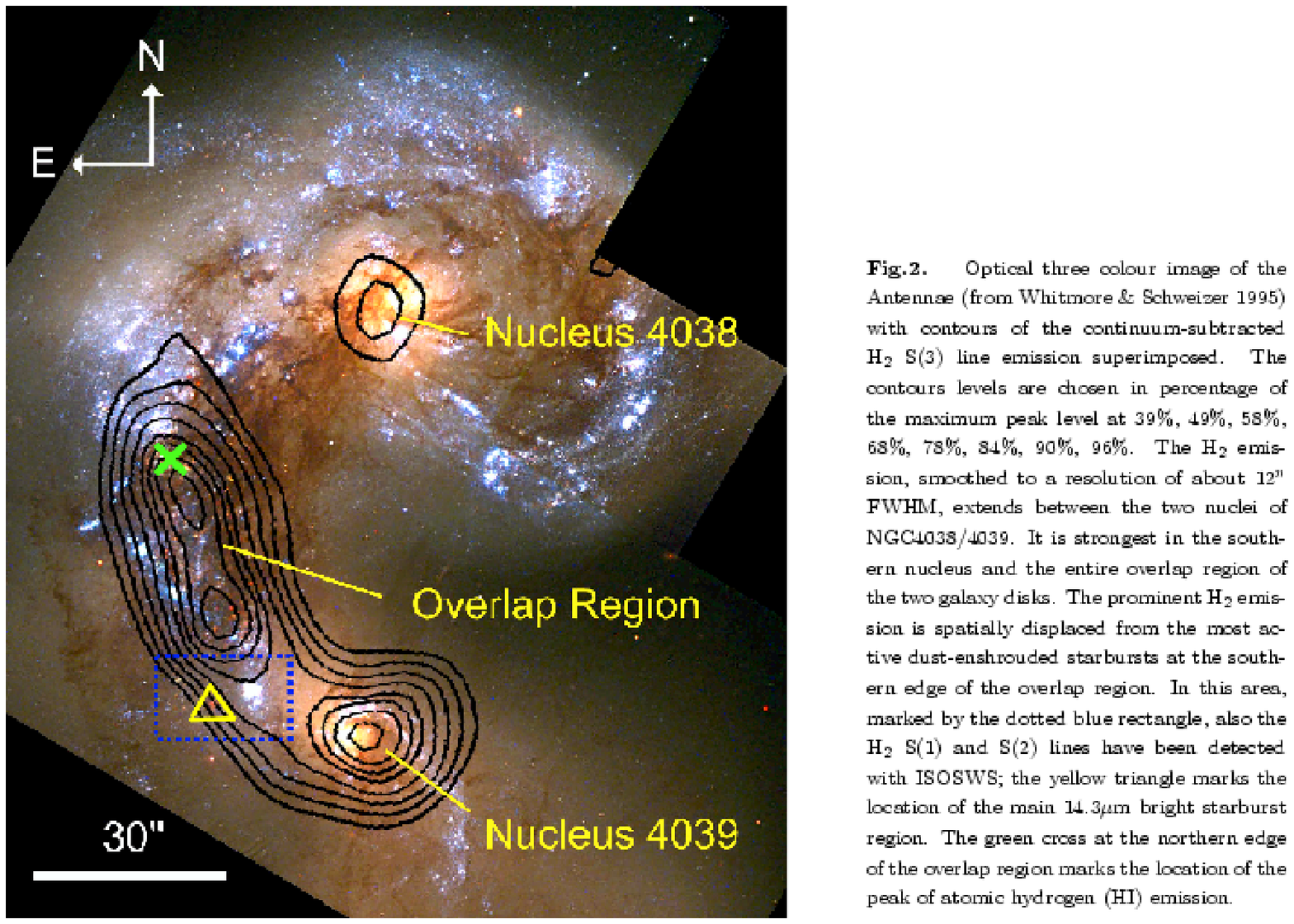 ,width=17cm, clip=true}
\end{figure*}
\addtocounter{figure}{+1}

The continuum-subtracted H$_{\rm 2}$  line flux integrated over the entire area
is 3.3$\times$10$^{\rm -15}$ Wm$^{\rm -2}$,
which corresponds to a line luminosity L(H$_{\rm 2}$) =
4.5$\times$10$^{\rm 7}$ L$_{\odot}$ for a distance of 21 Mpc; correcting for a screen
extinction A$_{\rm 9.7 \mu m}$ = 0.6 mag (Kunze et al. 1996)
this value rises to 8$\times$10$^{\rm 7}$ L$_{\odot}$. In
order to estimate how much of this H$_{\rm 2}$  luminosity can be attributed to
starbursts, we compare with the far-infrared luminosity
L$_{\rm FIR}$,  which is an ideal tracer for the power of ongoing and still
dust-enshrouded star formation events.
As shown in Figure 3, the
Antennae exhibit the highest L(H$_{\rm 2}$)/L$_{\rm FIR}$  ratio relative to other FIR
bright galaxies; their values were determined from the ISO Archive or
taken from the literature (Spoon et al. 2000, Rigopoulou et al. 2002, 
Lutz et al. 2003). While luminous and ultra-luminous
infrared galaxies in general have extinction corrected
L(H$_{\rm 2}$)/L$_{\rm FIR}$ 
ratios in the range of 10$^{\rm -5}$ to at most 10$^{\rm -4}$, the corresponding value
for the Antennae (1.25$\times$10$^{\rm -3}$) is more than ten times
higher. Remarkably, this value exceeds even that of NGC 6240, known as
the hitherto most pronounced H$_{\rm 2}$  emitter (Joseph et
al. 1984, Herbst et al. 1990, van der Werf et al. 1993).

The primary suggestion, that in NGC 6240 the
pre-starburst phase from the initial galaxy collision has been
detected, 
turned out to be questionable: Today the high L(H$_{\rm 2}$)/L$_{\rm FIR}$ 
ratio of this fairly advanced merger
is believed to stem from extreme supernova winds, setting in roughly 10
million years after the previous nuclear starbursts (Tecza et
al. 2000).
These superwinds
show up via the strong LINER spectrum and extended soft X-ray bubbles
and are running now against the gas between the two nuclei causing the
H$_{\rm 2}$ emission
(Ohyama et al. 2003, Max et al. 2005).

In contrast, in the rather virgin Antennae the exceptionally
high L(H$_{\rm 2}$)/L$_{\rm FIR}$  ratio is neither accompanied by a high total FIR
luminosity nor by a warm f$_{\rm 60 \mu m}$/f$_{\rm 100 \mu m}$
colour typical for most active
starbursts, nor by a conspicuous LINER-type spectrum (L{\'{\i}}pari et
al. 2003). Again, this
suggests that the H$_{\rm 2}$  emission cannot be a consequence of already
active starbursts: In case of UV fluorescence and X-ray excitation of
the H$_{\rm 2}$  line and even for shocks from supernova winds one would expect
that the accompanying starbursts are reflected by the far-infrared
luminosity, as it holds for luminous and ultra-luminous infrared
galaxies with L(H$_{\rm 2}$)/L$_{\rm FIR}$\,$<$\,10$^{\rm -4}$.

The exceptional  L(H$_{\rm 2}$)/L$_{\rm FIR}$ ratio
together with the spatial displacement of the H$_{\rm 2}$ 
emission from the known
vigorous starburst regions leads to the conclusion that in the
Antennae much of the molecular gas is in an extraordinary phase,
excited by a process not related to young stars and supernovae but
originating from pre-starburst shocks running through the
clouds. Since the shocks may quickly lead to the formation of massive
stars having a life time in the order of 10$^{\rm 6}$ years, the phase with
exceptionally high L(H$_{\rm 2}$)/L$_{\rm FIR}$
which we see for the Antennae might be
rather short-lived compared with the 10$^{\rm 8}$ years typically required for
the whole merger process. Hence it is observationally rare, suggesting
that such a pre-starburst phase is actually a common phenomenon during
the early encounter of colliding galaxies.

\begin{figure}
  \begin{center}
  \epsfig{file=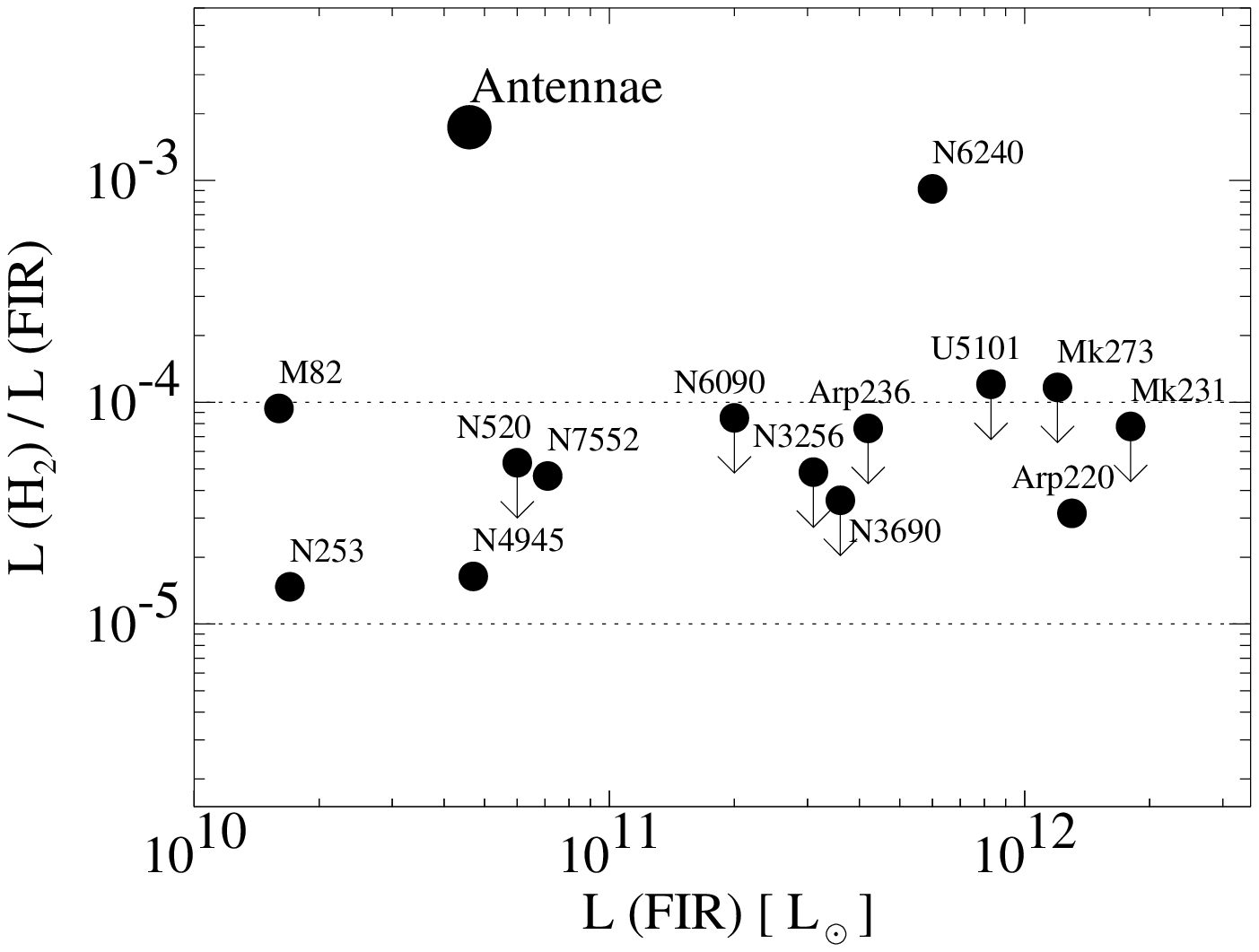 ,width=8.5cm, clip=true}
  \caption[]{ 
    Comparison of the ratio L(H$_{\rm 2}$  S(3))/L(FIR)  with the
    far-infrared 8 - 1000 $\mu$m luminosity. While
    the Antennae galaxy system has only a moderate
    FIR luminosity compared with other
    infrared galaxies, it exhibits the highest H$_{\rm 2}$  S(3) line to
    FIR luminosity ratio. The horizontal dotted lines indicate
    the range between 10$^{\rm -5}$
    and 10$^{\rm -4}$ found for infrared galaxies with
    obviously inconspicuous H$_{\rm 2}$  luminosity. 
    The L(H$_{\rm 2}$  S(3))/L$_{\rm FIR}$ 
    ratios of many sources are upper limits and, apart from the exotic merger
    N6240,  the extreme
    starburst galaxy M82  is the only object
    reaching L(H$_{\rm 2}$~S(3))/L$_{\rm FIR}$\,=\,10$^{\rm -4}$.
    Less active isolated
    galaxies like N253 and N4945 -- having 60 to 100 $\mu$m flux ratio
    below 0.5 -- show a ten times smaller L(H$_{\rm 2}$ S(3))/L$_{\rm
    FIR}$
    value of about
    10$^{\rm -5}$.
    For comparison, active starburst galaxies have f(60)/f(100)$\sim$1,
    while the Antennae have f(60)/f(100)$\sim$0.7.
    }
  \end{center}
\end{figure}

At a first glance, the shocks may arise from direct H$_{\rm 2}$  cloud-cloud
collisions due to the encounter of the two galaxy disks. A more
detailed consideration, however, suggests that essentially the atomic
HI clouds collide due to the higher impact efficiency, thereby
creating an overpressure medium which leads to shocks at the surfaces
of the H$_{\rm 2}$  clouds (Jog \& Solomon 1992).
In fact, high resolution (10") VLA maps of the Antennae
show enhanced HI emission at the northern edge of the overlap region
(Hibbard et al. 2001) marked by the green cross in
Figure 2, close to the area of the
bright H$_{\rm 2}$  line emission, in accordance with the overpressure model.

A rough estimate of the temperature and mass of the H$_{\rm 2}$ S(3)
emitting gas of the Antennae can be obtained by
standard procedures comparing with other H$_{\rm 2}$ line fluxes 
(c.f. Sect. 3.1. in  Rosenthal et al. 2000).
H$_{\rm 2}$ S(5) at $\lambda$=6.91$\mu$m is blended with
the [Ar II] $\lambda$=6.99$\mu$m line (Fig.\,1).
Adopting that about one sixth to one half of the observed 7$\mu$m
feature flux of
6\,$\times$\,10$^{\rm -15}$ Wm$^{-2}$  is due to H$_{\rm 2}$ S(5),
the temperature T$_{\rm S(5)-S(3)}$ is about 575-825\,K, a range also found
for 
other starburst galaxies (Rigopoulou et al. 2002, Lutz et al. 2003).
The mass of the molecular gas in the upper rotational level J=5 
is 2.6\,$\times$\,10$^{\rm 5}$ M$_{\odot}$.

The strength of the impending starbursts in the Antennae can be
estimated from the total amount of excited molecular gas, even at cooler 
temperatures emitting the H$_{\rm 2}$ v=0-0 S(1) line. 
H$_{\rm 2}$ S(1) line observations are only available for a
small 14"$\times$27" area in the southern overlap region. For
the entire Antennae we therefore extrapolate the H$_{\rm 2}$ S(1) line
luminosity from these ISOSWS observations
(Kunze et al. 1996). For this extrapolation we
adopt a constant H$_{\rm 2}$ S(1) / H$_{\rm 2}$ S(3) ratio across
the Antennae. Since this ratio might be higher in less active regions
outside the area covered by the ISOSWS observations, 
this is a conservative assumption leading to a lower limit on the actual
H$_{\rm 2}$ S(1) luminosity. We find that about 10\% of
the Antennae's total H$_{\rm 2}$ S(3) emission arise from the
14"$\times$27" area.  
If also the H$_{\rm 2}$  S(1) emission
from this ISOSWS area makes up 10\% of that in the 2'$\times$2' area,
the mass of excited H$_{\rm 2}$
observed in the S(1) transition at a temperature
T$\sim$200 K (adopted from Kunze et al. 2000)
yields about 4.9$\times$10$^{\rm 8}$ M$_{\odot}$ for
the entire Antennae; this corresponds to about 5\% of the total gas
mass of 9.6$\times$10$^{\rm 9}$ M$_{\odot}$ as derived from CO
observations covering a
comparable area (Young et al. 1995). An open issue is whether only the
currently
shocked gas is transformed into stars or whether much more of the gas
is involved in the future star formation process with only a small
fraction being currently gripped by the shock wave. Adopting that
about 20\% of the shocked H$_{\rm 2}$  of
5$\times$10$^{\rm 8}$ M$_{\odot}$ collapses into stars during
the next 10$^{\rm 6}$ years, the resulting average star forming rate is 100
M$_{\odot}$/year. This is an extreme value found only for ultra-luminous
infrared galaxies. If the gas mass of
10$^{\rm 8}$ M$_{\odot}$  is converted to stars of
about 20 M$_{\odot}$, each having a luminosity of about 10$^{\rm 4}$
L$_{\odot}$, the luminosity
of the Antennae will increase by  about 5$\times$10$^{\rm 10}$
L$_{\odot}$, hence it will be
doubled. In order to become ultra-luminous much heavier nuclear
starbursts are required during further stages of the merger process.

Eventually, the shock-induced star formation may have played a role in
the evolution of proto-galaxies in the early universe at a
redshift z$\sim$20. The very first stellar generation with zero
metallicity, the population III stars, must have formed during a short
episode; otherwise their metallicity would have been raised. Due to
the lack of appropriate radiative cooling via metals the kinetic
energy in the gas clouds remains at a high level. Therefore, the
clouds need time to accumulate the higher amount of gas required
before they can collapse according to the Jeans criterion. In order to
fit the constraints, strong simplifications for the processes in the
early universe have to be made (e.g. Abel et al. 2002).  Our results
for the Antennae
suggest that shocks could provide a natural trigger for speeding-up
the collapse of clouds also in colliding proto-galaxies.

\acknowledgements 
It is a pleasure for us to thank Olivier Laurent for help with the
    ISOCAM-CVF data reduction, Rainer Beck, Hans Hippelein and
    Theodor Schmidt-Kaler for stimulating discussions,
    and the referee Bruno Altieri for constructive suggestions. 
    For photometry NED and SIMBAD
    were used. This  research was supported by 
    Nordrhein-Westf\"alische Akademie der Wissenschaften
    and by Deutsches Zentrum f\"ur Luft- und Raumfahrt (DLR).

\end{document}